\documentclass[10pt]{article}
\usepackage{graphicx}
\usepackage{amsmath}
\usepackage{amssymb}
\usepackage{caption2}
\begin{document}
\bibliographystyle{prsty}
\begin{center}
{\large {\bf \sc{ Analysis of the radiative decays among the bottomonium states  }}} \\[2mm]
Zhi-Gang Wang  \footnote{E-mail:wangzgyiti@yahoo.com.cn. } \\
  Department of Physics, North China Electric Power University, Baoding 071003, P. R.
  China
\end{center}

\begin{abstract}
In this article, we perform an systematic study of  the radiative
transitions among the bottomonium states using  the
 heavy quarkonium effective Lagrangians, and make  predictions for the ratios among the
radiative decay widths of a special multiplet to another multiplet.
The predictions can be confronted with the experimental data in the
future.
\end{abstract}

PACS numbers:  14.40.Pq; 13.40.Hq

{\bf{Key Words:}}  Bottomonium states,  Radiative decays
\section{Introduction}

In recent years, the Babar, Belle, CLEO, CDF, D0 and FOCUS
collaborations have discovered (or confirmed) a large number of
charmonium-like states
  and revitalized  the interest
in the spectroscopy of the charmonium states
\cite{Olsen-XYZ,ReviewEichten,ReviewCharm,Recent-review}. There are
also some progresses  in the spectroscopy of the bottomonium states.
In 2004, the CLEO collaboration observed the $\Upsilon(1{\rm D})$
states in the four-photon decay cascade, $\Upsilon(3{\rm S})\to
\gamma \chi_b(2{\rm P})$, $\chi_b(2{\rm P})\to \gamma \Upsilon(1{\rm
D})$, $\Upsilon(1{\rm D})\to \gamma \chi_b(1{\rm P})$, $\chi_b(1{\rm
P})\to \gamma \Upsilon(1 {\rm S})$, and obtained  the mass
$M_{\Upsilon(1^3{\rm D}_2)}=(10161.1\pm0.6\pm1.6) \,\rm{MeV}$
\cite{CLEO2004}.   In 2008, the Babar collaboration observed the
$\eta_b(1 {\rm S})$ in the radiative decay $\Upsilon(3{\rm S}) \to
\gamma \eta_b(1{\rm S})$ \cite{Babar2008}, and latter confirmed it
in the radiative decay $\Upsilon(2{\rm S}) \to \gamma \eta_b(1{\rm
S})$ \cite{Babar0903}.
 In 2010, the Babar collaboration observed the
$\Upsilon(1^3{\rm D}_j)$ state through the decay chain
 $\Upsilon(3{\rm S})\to\gamma\gamma\Upsilon(1^3{\rm
D}_j)\to\gamma\gamma\pi^+\pi^-\Upsilon(1{\rm S})$ with $j=1,2,3$, and obtained the
mass  $M_{\Upsilon(1^3{\rm D}_2)}=(10164.5\pm0.8 \pm0.5)\,\rm{MeV} $
\cite{Babar1004}.  In 2011, the Belle collaboration reported the
observation of the  spin-singlet
bottomonium states $h_b(1\rm{P})$ and $h_b(2\rm{P})$, which are produced in the reactions  $e^+e^- \to
h_b({\rm{nP}})\pi^+\pi^-$ with significances of $5.5\,\sigma$ and
$11.2\,\sigma$, respectively \cite{Belle1103}. The measured masses
are $M_{h_b(1{\rm
{P}})}=\left(9898.25\pm1.06^{+1.03}_{-1.07}\right)\,\rm{MeV}$ and
$M_{h_b(2{\rm{P}})}=\left(10259.76\pm0.64^{+1.43}_{-1.03}\right)\,\rm{MeV}$,
respectively.   Recently, the ATLAS collaboration observed the $\chi_{bj}(\rm{3P})$ with $j=1,2$ in the radiative  transitions  $\chi_{bj}(\rm{3P})\to \gamma \Upsilon(\rm{1S})$,
  $ \gamma \Upsilon(\rm{2S})$ in the proton-proton collisions at the Large Hadron Collider (LHC) at the energy $\sqrt{s} = 7\,\rm{TeV}$ \cite{ATLAS2011}. The measured mass barycenter is $(10530 \pm 5 \pm 9)\, \rm{MeV}$, and the hyperfine mass splitting is fixed
to the theoretically predicted value of $12\,\rm{ MeV}$.
    And more bottomonium states would be observed in the
future at the Tevatron, KEK-B, RHIC and LHCb.

On the other hand, there have been several theoretical works on the
spectroscopy of the bottomonium states, such as the relativized
potential model (Godfrey-Isgur model) \cite{Godfrey1985}, the
Cornell potential model, the logarithmic potential model,
  the power-law potential model, the QCD-motivated potential model
\cite{EQ1994}, the  relativistic quark model based on a
quasipotential approach in QCD \cite{Ebert2003},  the Cornell
potential model combined with heavy quark mass expansion
\cite{Roberts1995}, the screened potential model \cite{LiChao0909},
the potential non-relativistic QCD model \cite{pNRQCD}, the
confining  potential model with the Bethe-Salpeter equation
\cite{BSE},  etc. In Table 1, we list the experimental values of the
bottomonium states  compared with some theoretical predictions
\cite{Belle1103,ATLAS2011,Godfrey1985,LiChao0909,PDG}.

The charmonium and bottomonium states have analogous properties, the
hadronic transitions and radiative transitions among the heavy
quarkonium states  have been studied by the QCD multipole expansion
\cite{Kuang06}, the nonrelativistic potential model
\cite{EM-1,EM-2,EM-3,E1-form}, the heavy quarkonium  effective theory
\cite{PRT1997}, the coupled-channel approach \cite{EM-2,EM-3}, the
hybrid approach based on the multipole expansion and heavy quark
symmetry \cite{ChoWise}, etc. In   the nonrelativistic
potential models,  the $E_1$ and $M_1$  transitions among the bottomonium states
 are usually studied by the following formulae    \cite{EM-1,EM-2,EM-3,E1-form},
\begin{eqnarray}
 \Gamma_{E_1}\left({ n}^{2s+1}L_j \to{{ n}^\prime}^{2s'+1}{L'}_{j'}+\gamma\right)&=&\frac{4}{3}e_b^2\alpha
 E^3_\gamma \frac{E_f}{M_i}\delta_{ss'}C_{fi}\mid \langle {{ n}^\prime}^{2s'+1}{L'}_{j'}\mid r\mid { n}^{2s+1}L_j
 \rangle\mid^2 \, , \nonumber \\
\Gamma_{M_1}\left({ n}^{2s+1}L_j \to{{n}^\prime}^{2s'+1}{L'}_{j'}+\gamma\right)&=&\frac{4}{3}e_b^2\frac{\alpha}{m_b^2}
 E^3_\gamma \frac{E_f}{M_i}\frac{2j'+1}{2L+1}\delta_{LL'}\delta_{ss'\pm1}\mid \langle {{ n}^\prime}^{2s'+1}{L'}_{j'}\mid  { n}^{2s+1}L_j
 \rangle\mid^2 \, ,\nonumber \\
\end{eqnarray}
where the $E_{\gamma}$ is the photon energy, the $E_f$ is the energy
of the final  bottomonium state, the $ M_i$ is the mass of the
initial  bottomonium state, and the angular matrix factor $C_{fi}$
is
\begin{equation}
C_{fi}={\rm max}(L,  L')(2j'+1)\left\{
\begin{array}{ccc}
 L'&  J' &s\\
 J &  L  &1
\end{array}\right\}^2 \, .
\end{equation}
The values of the matrix elements  $\langle {{n}^\prime}^{2s'+1}{L'}_{j'}\mid r\mid { n}^{2s+1}L_j
 \rangle$ and $\langle {{n}^\prime}^{2s'+1}{L'}_{j'}\mid { n}^{2s+1}L_j
 \rangle$ and their $v^2/c^2$ corrections depend on the details of the wave-functions,  which are evaluated
  using a special potential model \cite{Godfrey1985,EQ1994,Ebert2003,LiChao0909}.
In Ref.\cite{Wang1101}, we focus on
the traditional charmonium scenario of the new charmonium-like
states and study the radiative transitions among the charmonium
states with the heavy quarkonium (or meson)  effective theory based on the heavy quark symmetry
\cite{PRT1997,RevWise,RevNeubert}, which have been applied to
identify the excited $D_s$ and $D$ mesons, such as the $D_s(3040)$,
$D_s(2700)$, $D_s(2860)$, $D(2550)$, $D(2600)$, $D(2750)$ and
$D(2760)$
\cite{Colangelo1001,Colangelo0710,Colangelo0607,Colangelo0511,Wang1009}.
In this article, we extend our previous works to study the radiative
transitions among the bottomonium states using the heavy quarkonium
effective theory.

\begin{table}
\begin{center}
\begin{tabular}{|cc|c|c|c|}\hline
 \multicolumn{2}{|c|}{States} & Experimental data \cite{Belle1103,ATLAS2011,PDG} & Theoretical values \cite{LiChao0909}  & Theoretical values \cite{Godfrey1985}\\ \hline
1S &  $\Upsilon(1^3{\rm S}_1)$    & $ 9460.30\pm 0.26$                       & 9460                    & 9460    \\
   &  $\eta_b(1^1{\rm S}_0)$      & $ 9390.9\pm 2.8 $                        & 9389                    & 9400    \\ \hline
2S &  $\Upsilon(2^3{\rm S}_1)$    & $ 10023.26\pm 0.31$                      & 10016                   & 10000   \\
   &  $\eta_b'(2^1{\rm S}_0)$     &                                          & 9987                    & 9980    \\
\hline
3S &  $\Upsilon(3^3{\rm S}_1)$    & $ 10355.2\pm 0.5$                        & 10351                   & 10350   \\
   &  $\eta_b(3^1{\rm S}_0)$      &                                          & 10330                   & 10340   \\
\hline
4S &  $\Upsilon(4^3{\rm S}_1)$    & $ 10579.4\pm 1.2$                        & 10611                   & 10630   \\
   &  $\eta_b(4^1{\rm S}_0)$      &                                          & 10595                   &        \\
\hline
5S &  $\Upsilon(5^3{\rm S}_1)$    & $ 10865\pm 8$                            & 10831                   & 10880   \\
   &  $\eta_b(5^1{\rm S}_0)$      &                                          & 10817                   &        \\
\hline
6S &  $\Upsilon(6^3{\rm S}_1)$    & $ 11019\pm 8$                            & 11023                   & 11100   \\
   &  $\eta_b(6^1{\rm S}_0)$      &                                          & 11011                   &        \\
\hline
7S &  $\Upsilon(7^3{\rm S}_1)$    &                                          & 11193                   &        \\
   &  $\eta_b(7^1{\rm S}_0)$      &                                          & 11183                   &        \\
\hline
1P &  $\chi_{b2}(1^3{\rm P}_2)$   & $ 9912.21\pm 0.26\pm 0.31$               & 9918                    & 9900    \\
   &  $\chi_{b1}(1^3{\rm P}_1 )$  & $ 9892.78 \pm 0.26\pm 0.31$              & 9897                    & 9880    \\
   &  $\chi_{b0}(1^3{\rm P}_0)$   & $ 9859.44 \pm 0.42\pm 0.31$              & 9865                    & 9850    \\
   &  $h_b(1^1{\rm P}_1)$         & $ 9898.25 \pm 1.06^{+1.03}_{-1.07}$      & 9903                    & 9880    \\
\hline
2P &  $\chi_{b2}(2^3{\rm P}_2)$   & $10268.65\pm0.22\pm0.50$                 & 10269                   & 10260   \\
   &  $\chi_{b1}(2^3{\rm P}_1)$   & $10255.46 \pm0.22\pm0.50$                & 10251                   & 10250   \\
   &  $\chi_{b0}(2^3{\rm P}_0)$   & $10232.5\pm0.4\pm0.5$                    & 10226                   & 10230   \\
   &  $h_b(2^1{\rm P}_1) $        & $10259.76 \pm 0.64^{+1.43}_{-1.03}$      & 10256                   & 10250   \\
\hline
3P &  $\chi_{b2}(3^3{\rm P}_2)$   & $10536\pm 5 \pm 9$                       & 10540                   &        \\
   &  $\chi_{b1}(3^3{\rm P}_1)$   & $10524\pm 5 \pm 9$                       & 10524                   &        \\
   &  $\chi_{b0}(3^3{\rm P}_0)$   &                                          & 10502                   &        \\
   &  $h_b(3^1{\rm P}_1)$         &                                          & 10529                   &        \\
\hline
4P &  $\chi_{b2}(4^3{\rm P}_2)$   &                                          & 10767                   &        \\
   &  $\chi_{b1}(4^3{\rm P}_1)$   &                                          & 10753                   &        \\
   &  $\chi_{b0}(4^3{\rm P}_0)$   &                                          & 10732                   &        \\
   &  $h_b(4^1{\rm P}_1)$         &                                          & 10757                   &        \\
\hline
5P &  $\chi_{b2}(5^3{\rm P}_2)$   &                                          & 10965                   &        \\
   &  $\chi_{b1}(5^3{\rm P}_1)$   &                                          & 10951                   &        \\
   &  $\chi_{b0}(5^3{\rm P}_0)$   &                                          & 10933                   &        \\
   &  $h_b(5^1{\rm P}_1)$         &                                          & 10955                   &        \\
\hline
1D &  $\Upsilon_3(1^3{\rm D}_3)$  &                                          & 10156                   & 10160   \\
   &  $\Upsilon_2(1^3{\rm D}_2)$  & $10161\pm0.6\pm1.6$                      & 10151                   & 10150   \\
   &  $\Upsilon(1^3{\rm D}_1) $   &                                          & 10145                   & 10140   \\
   &$\eta_{b2}(1^1{\rm D}_2)$     &                                          & 10152                   & 10150   \\
\hline
2D &  $\Upsilon_3(2^3{\rm D}_3)$  &                                          & 10442                   & 10450   \\
   &  $\Upsilon_2(2^3{\rm D}_2)$  &                                          & 10438                   & 10450   \\
   &  $\Upsilon(2^3{\rm D}_1)$    &                                          & 10432                   & 10440   \\
   &$\eta_{b2}(2^1{\rm D}_2)$     &                                          & 10439                   & 10450   \\
\hline
3D &  $\Upsilon_3(3^3{\rm D}_3)$  &                                          & 10680                   &        \\
   &  $\Upsilon_2(3^3{\rm D}_2)$  &                                          & 10676                   &        \\
   &  $\Upsilon(3^3{\rm D}_1) $   &                                          & 10670                   &        \\
   &$\eta_{b2}(3^1{\rm D}_2)$     &                                          & 10677                   &        \\
\hline
4D &  $\Upsilon_3(4^3{\rm D}_3) $ &                                          & 10886                   &        \\
   &  $\Upsilon_2(4^3{\rm D}_2)$  &                                          & 10882                   &        \\
   &  $\Upsilon(4^3{\rm D}_1)$    &                                          & 10877                   &        \\
   &$\eta_{b2}(4^1{\rm D}_2)$     &                                          & 10883                   &        \\
\hline
5D &  $\Upsilon_3(5^3{\rm D}_3)$  &                                          & 11069                   &        \\
   &  $\Upsilon_2(5^3{\rm D}_2)$  &                                          & 11065                   &        \\
   &  $\Upsilon(5^3{\rm D}_1)$    &                                          & 11060                   &        \\
   &$\eta_{b2}(5^1{\rm D}_2)$     &                                          & 11066                   &        \\
\hline \hline
\end{tabular}
\caption{Experimental and theoretical mass spectrum of the
bottomonium states, where the unit is MeV. }
\end{center}
\end{table}

The article is arranged as follows:  we study  the   radiative
transitions  among the bottomonium  states with the heavy quarkonium effective Lagrangians   in Sect.2; in Sect.3, we present the
 numerical results and discussions; and Sect.4 is reserved for our
conclusions.

\section{ The radiative transitions with the heavy quarkonium effective Lagrangians }
In the infinite heavy quark mass limit, the heavy quarkonium states do not have  heavy quark flavor symmetry and  spin symmetry, while for the intermediate heavy quark mass, the heavy quark spin symmetry is expected to  make sense \cite{GattoPLB92,GattoPLB93}. In fact, the $c$ and $b$ quarks have large but finite masses, we can construct heavy quarkonium effective Lagrangians based on the heavy quark spin symmetry.
In the nonrelativistic QCD for the heavy quark systems, we introduce three typical energy scales $m_Q$, $m_Qv$, $m_Qv^2$, where $m_Q$ and $v$ are the heavy quark masses and velocities respectively, and count the operators with the  power orders of $v$,  take the
 heavy quarkonia  as bound states and study them with the nonrelativistic Schrodinger field theory, and apparent Lorentz covariance is lost. In the heavy quark effective theories,  we introduce two typical energy scales $m_Q$ and $\Lambda_{QCD}$, count the operators with the  power orders of $1/m_Q$,  and take the heavy quarkonia  as the basic relativistic quantum fields rather than   bound states at the hadron level, the calculations are more simple, and apparent Lorentz covariance is maintained. In the two approaches, the heavy quarkonium states are classified in the same scheme. And the two approaches both have advantages and shortcomings.

The  bottomonium states can be classified according to the
 notation  $n^{2s+1}L_{j}$, where the $n$ is the radial quantum number, the $L$ is
  the orbital angular momentum, the $s$ is  the spin, and the $j$ is the total angular
momentum. They have the parity and charge conjugation $P=(-1)^{L+1}$
and $C=(-1)^{L+s}$, respectively. The states have the same radial
quantum number $ n$ and  orbital momentum $L$ can be expressed
by  the superfields $J$, $J^\mu$, $J^{\mu\nu}$, etc
\cite{GattoPLB92,GattoPLB93},
\begin{eqnarray}
J&=&\frac{1+{\rlap{v}/}}{2}\left\{\Upsilon_{\mu}\gamma^\mu-\eta_b\gamma_5\right\}
\frac{1-{\rlap{v}/}}{2} \, , \nonumber \\
J^\mu&=&\frac{1+{\rlap{v}/}}{2}\left\{\chi_{2}^{\mu\nu}\gamma_\nu+\frac{1}{\sqrt{2}}\epsilon^{\mu\alpha\beta\lambda}v_\alpha
\gamma_{\beta}\chi^{1}_{\lambda}+\frac{1}{\sqrt{3}}\left(\gamma^\mu-v^\mu\right)\chi_{0}+h^\mu_b\gamma_5\right\}
\frac{1-{\rlap{v}/}}{2} \, , \nonumber \\
J^{\mu\nu}&=&\frac{1+{\rlap{v}/}}{2}\left\{\Upsilon_3^{\mu\nu\alpha}\gamma_\alpha+\frac{1}{\sqrt{6}}
\left[\epsilon^{\mu\alpha\beta\lambda}v_\alpha
\gamma_{\beta}g^{\tau\nu}+\epsilon^{\nu\alpha\beta\lambda}v_\alpha
\gamma_{\beta}g^{\tau\mu}\right]\Upsilon^2_{\tau\lambda}+\right.\nonumber\\
&&\left.\left[\sqrt{\frac{3}{20}}\left[\left(\gamma^\mu-v^\mu\right)g^{\nu\alpha}+\left(\gamma^\nu-v^\nu\right)g^{\mu\alpha}\right]-\frac{1}{\sqrt{15}}\left(g^{\mu\nu}-v^{\mu}v^{\nu}\right)\gamma^\alpha\right]\Upsilon_\alpha+\eta^{\mu\nu}_{2}\gamma_5\right\}
\frac{1-{\rlap{v}/}}{2} \, , \nonumber\\
\end{eqnarray}
where the $v^{\mu}$ denotes the four velocity associated to the
superfields. We multiply  the  bottomonium fields
$\Upsilon^3_{\mu\nu\alpha}$, $\Upsilon^2_{\mu\nu}$,
$\Upsilon_{\mu}$, $\eta^2_{\mu\nu}$, $\chi^2_{\mu\nu}$, $\cdots$
with the factors $\sqrt{M_{\Upsilon_3}}$, $\sqrt{M_{\Upsilon_2}}$,
$\sqrt{M_{\Upsilon}}$, $\sqrt{M_{\eta_2}}$, $\sqrt{M_{\chi_2}}$,
$\cdots$, respectively, and they have dimension of mass $\frac{3}{2}$. The
superfields $J$, $J^\mu$, $J^{\mu\nu}$ are functions of the radial
quantum numbers $ n$, the fields in a definite superfield have
the same $ n$, and form a multiplet. The superfields
$J^{\mu_1\ldots\mu_L}$ have the following properties under the
parity, charge conjunction, heavy quark spin transformations,
\begin{eqnarray}
J^{\mu_1\ldots\mu_L}&\stackrel{P}{\longrightarrow}&\gamma^{0}J_{\mu_1\ldots\mu_L}\gamma^{0} \, ,  \nonumber \\
J^{\mu_1\ldots\mu_L}&\stackrel{C}{\longrightarrow}&(-1)^{L+1}C[J_{\mu_1\ldots\mu_L}]^{T}C \,,\nonumber \\
J^{\mu_1\ldots\mu_L}&\stackrel{S}{\longrightarrow}&SJ_{\mu_1\ldots\mu_L}S^{\prime\dagger}\,,\nonumber \\
v^{\mu}&\stackrel{P}{\longrightarrow}&v_{\mu}\, ,
\end{eqnarray}
where $S,S^{\prime}\in SU(2)$ heavy quark spin symmetry groups, and
$[S,{\rlap{v}/}]=[S^{\prime},{\rlap{v}/}]=0$. The   $S$-wave   multiplet   contains  the spin-singlet bottomonium states; while
the $P$-wave and $D$-wave  multiplets  contain  both the spin-singlet and spin-triplet bottomonium states, there exist mass splittings among (or between)  the bottomonium states in the same multiplet.
In the nonrelativistic  quark models, we resort to the spin-spin, spin-orbit and tensor interactions to take into account the mass splittings.  In the present case, we can  use the superfields $J$, $J^\mu$, $J^{\mu\nu}$ and  the Dirac matrix $\sigma_{\mu\nu}$ to construct the heavy quarkonium effective Lagrangians at the hadronic level to account for the mass-splittings in a multiplet \cite{GattoPLB93}.
The heavy quark effective Lagrangian  can be written as
\begin{eqnarray}
  {\cal L}  &=&\bar h_viv\cdot Dh_v+{1\over2m_Q}  \left[\bar h_v(iD_\perp)^2h_v
  +{g\over2}\,\bar h_v\sigma^{\alpha\beta}  G_{\alpha\beta}h_v\right]  +\dots\,,
\end{eqnarray}
where $D^\mu_\perp=D^\mu-v^\mu v\cdot D$, the $D_\mu$ is the
covariant derivative, and the $h_v$ is the heavy quark field.
The heavy quark spin symmetry breaking terms are of the order  $\mathcal{O}(1/m_Q)$, $\mathcal{O}(1/m_Q^2)$, $\cdots$, and the corresponding corrections to the heavy quarkonium states (or mesons)  can be counted as  $\Lambda_{QCD}/m_Q$, $(\Lambda_{QCD}/m_Q)^2$, $\cdots$, respectively, here we introduce the scale $\Lambda_{QCD}$
to characterize   the bound states. The mass-splittings in a multiplet can also be reproduced with the heavy quarkonium effective Lagrangians at the next-to-leading order \cite{GattoPLB93}.

The radiative transitions between the $ m$ and $ n$
bottomonium states  can be described by the following heavy quarkonium effective
Lagrangians \cite{PRT1997,Wang1101,Fazio2008}\footnote{The
Lagrangian
 ${\cal{L}}_{PD}$ was introduced by F. De Fazio in Ref.\cite{Fazio2008} for the first time, I failed to
  take notice of the article, and constructed the Lagrangian
  ${\cal{L}}_{PD}$ independently in Ref.\cite{Wang1101},
  furthermore, I constructed spin violation  Lagrangian
  ${\cal{L}}_{SS}$ in Ref.\cite{Wang1101}. When the present  article was submitted to http://arxiv.org and appeared as arXiv:1101.2506,
 Dr. F. De Fazio drew my  attention to Ref.\cite{Fazio2008}.  },
\begin{eqnarray}
{\cal{L}}_{SS}&=&\sum_{m,n}\delta(m,n) \mathrm{Tr}\left[\bar{J}(m)\sigma_{\mu\nu}J(n)\right]F^{\mu\nu} \, , \nonumber\\
{\cal{L}}_{SP}&=&\sum_{m,n}\delta(m,n)\mathrm{Tr}\left[\bar{J}({m})J_{\mu}(n)+\bar{J}_\mu(n)
J(m)\right]V^{\mu} \, , \nonumber\\
{\cal{L}}_{PD}&=&\sum_{m,n}\delta(m,n)\mathrm{Tr}\left[\bar{J}_{\mu\nu}({m})J^{\nu}(n)+\bar{J}^\nu(n)
J_{\mu\nu}(m)\right]V^{\mu} \, ,
\end{eqnarray}
where $\bar{J}_{\mu_1\ldots\mu_L}=\gamma^0
J_{\mu_1\ldots\mu_L}^{\dag} \gamma^0$, $V^\mu=F^{\mu\nu}v_\nu$, the
$F^{\mu\nu}$ is the electromagnetic field tensor, and the $\delta(m,n)$ are
the coupling constants and have different values in the ${\cal{L}}_{SS}$, ${\cal{L}}_{SP}$ and ${\cal{L}}_{PD}$.

The Lagrangians ${\cal{L}}_{SP}$ and
${\cal{L}}_{PD}$ preserve parity, charge conjugation,    gauge
invariance and heavy quark spin symmetry, while the Lagrangian
${\cal{L}}_{SS}$ violates the heavy quark spin symmetry.
 The effective Lagrangians ${\cal{L}}_{SP}$ and ${\cal{L}}_{PD}$
describing the electric dipole $E_1$ transitions can be realized in
the leading order $\mathcal {O}(1)$ in the heavy quark effective theory, while the
 effective Lagrangian ${\cal{L}}_{SS}$ describing the
magnetic dipole $M_1$ transitions can be realized in the
next-to-leading order $\mathcal {O}(1/m_Q)$. The corrections to the ${\cal{L}}_{SP}$,
${\cal{L}}_{PD}$ and ${\cal{L}}_{SS}$  come from  the
next-to-leading order $\mathcal {O}(1/m_Q)$ and the
next-to-next-to-leading order $\mathcal {O}(1/m^2_Q)$, respectively. We can construct the corresponding Lagrangians
with the  superfields $J$, $J^\mu$, $J^{\mu\nu}$,  the Dirac matrix $\sigma_{\mu\nu}$, the electromagnetic
field tensor $F_{\mu\nu}$ and the four-vector $v_\mu$, and introduce additional unknown
coupling constants, which have to be fitted to the precise experimental data in the future, and study the
spin symmetry violations   in the radiative decays to the bottomonium states in an special multiplet.

In the case of the charmonium states,  the $\mathcal {O}(v^2/c^2)$ corrections to the $E_1$ transitions come from the magnetic quadrupole $M_2$ and
 electric octupole $E_3$ terms,  the current average values of the relative
amplitudes are  $M_2/\sqrt{E_1^2+M_2^2+E_3^2} =(-10.0\pm 1.5)\times 10^{-2}$ and $ E_3/\sqrt{E_1^2+M_2^2+E_3^2} =(1.6\pm 1.3)\times 10^{-2}$ in the $\chi_{c2}(1{\rm P})\to \gamma {J/\psi}$ decay, $M_2/\sqrt{E_1^2+M_2^2+E_3^2} =(1.0\pm 1.4)\times 10^{-2}$ and $ E_3/\sqrt{E_1^2+M_2^2+E_3^2} =(-1.0\pm 1.1)\times 10^{-2}$ in the $ \psi(2{\rm S})\to \gamma \chi_{c2}(1{\rm P})$ decay,
$M_2/\sqrt{E_1^2+M_2^2+E_3^2} =(-5.4^{+1.2}_{-1.5})\times 10^{-2}$  in the $\chi_{c1}(1{\rm P})\to \gamma {J/\psi}$ decay, $M_2/\sqrt{E_1^2+M_2^2+E_3^2} =(2.9\pm 0.8)\times 10^{-2}$  in the $ \psi(2{\rm S})\to \gamma \chi_{c1}(1{\rm P})$ decay from the Particle Data Group \cite{PDG}. Although the values differ from the theoretical expectations  $v^2/c^2\approx 0.3$, the corrections are very small, we expect that the corresponding corrections for the bottomonium states are also very small.

In the screened potential model \cite{LiChao0909},
the predictions for the decay widths  of the transitions $\Upsilon(2{\rm S}) \to \chi_{bj}(1{\rm P}) \gamma$, $\Upsilon(3{\rm S}) \to \chi_{bj}(2{\rm P}) \gamma$, $j=0,1,2$,  become   better compared to the experimental data after the $\mathcal {O}(v^2/c^2)$ corrections are taken into account, the $\mathcal {O}(v^2/c^2)$ corrections  are not large; on the other hand, the $\mathcal {O}(v^2/c^2)$ corrections in the radiative transitions $4{\rm S} \to 2{\rm P}$, $4{\rm S} \to 1{\rm P}$, $\cdots$, are very large, and it is not a good approximation for taking the spin symmetry breaking effects perturbatively.  At the hadron  level, large spin symmetry violations mean that the   $\mathcal {O}(1/m_Q)$  corrections are large enough to ruin the leading order approximation, while we do not have enough experimental data
 to fit the unknown coupling constants in the phenomenological Lagrangians, and those parameters cannot be canceled out with each other to result in independent  ratios among the radiative decay widths; we expect that neglecting the next-to-leading order and next-to-next-to-leading order corrections for the ${\cal{L}}_{SP}$, ${\cal{L}}_{PD}$ and ${\cal{L}}_{SS}$ respectively  leads to   uncertainties of the order $\mathcal {O}(\Lambda_{QCD}/m_b)$;
in other words, we expect that the flavor and spin symmetry breaking
corrections of the order $\mathcal {O}(1/m_Q)$ to the effective
Lagrangians ${\cal{L}}_{SP}$ and ${\cal{L}}_{PD}$ are  smaller than
(or not as large as) the leading order contributions.

 In the heavy quark
limit, the contributions of the order $\mathcal {O}(1/m_Q)$ are
greatly suppressed.  For example, the branching ratios of the
radiative decay widths of the $\Upsilon(2{\rm{S}})$ have the
hierarchy ${\rm Br}\left(\Upsilon(2{\rm{S}})\to
\chi_{bj}(1{\rm{P}})\gamma\right)\sim 10^{-2}\gg{\rm
Br}\left(\Upsilon(2{\rm{S}})\to \eta_{b}(1{\rm{S}})\gamma\right)\sim
10^{-4}$, $j=0,1,2$ \cite{PDG}, although the radiative decays
$\Upsilon(2{\rm{S}})\to \chi_{bj}(1{\rm{P}})\gamma$ are suppressed
in the phase-space. In the case of the charmonium states, the decay widths
 $ \Gamma(\psi^{\prime}\to\chi_{c0}\gamma)=29.2\,\rm{KeV}$,
 $\Gamma(\psi^{\prime}\to\chi_{c1}\gamma)=28.0\,\rm{KeV}$,
 $\Gamma(\psi^{\prime}\to\chi_{c2}\gamma)=26.6\,\rm{KeV}$,
 $\Gamma(\chi_{c0}\to J/\psi\gamma)=119.5\,\rm{KeV}$,
$ \Gamma(\chi_{c1}\to J/\psi\gamma)=295.8\,\rm{KeV}$,
$ \Gamma(\chi_{c2}\to J/\psi\gamma)=384.2\,\rm{KeV}$ in the $E_1$ transitions are very large compared with
  the decay widths $\Gamma(\psi'\to \gamma\eta_c)= 1.03\,\rm{KeV}$  and $\Gamma(J/\psi \to \gamma\eta_c)= 1.58\,\rm{KeV}$ in the $M_1$ transitions \cite{PDG}, as the $M_1$ transitions take place at the order $\mathcal O(1/m_Q)$.  However, it is difficult to count for that the
decay widths $\Gamma(\psi'\to \gamma\eta_c)= 1.03\,\rm{KeV}$  and $\Gamma(J/\psi \to \gamma\eta_c)= 1.58\,\rm{KeV}$ are of the same order   by the
$\mathcal O(1/m_Q)$ depression. In the nonrelativistic potential models, the overlap matrix elements $\langle {{ n}^\prime}^{2s'+1}{S'}_{j'} \mid {n}^{2s+1}S_j
 \rangle$ in the leading order approximation are $1$ and $0$ for the transitions $J/\psi \to \gamma\eta_c$ and $\psi'\to \gamma\eta_c$ respectively, the perturbative
 $\alpha_s$ corrections and relativistic $v^2/c^2$ corrections cannot count for  the
decay widths $\Gamma(\psi'\to \gamma\eta_c)= 1.03\,\rm{KeV}$  and $\Gamma(J/\psi \to \gamma\eta_c)= 1.58\,\rm{KeV}$ simultaneously \cite{PDG}, we have to suppose large high order corrections or introduce new mechanisms  \cite{ReviewCharm,Brambilla2011}. In the heavy quarkonium effective theory, the heavy quark spin symmetry cannot count for the analogous   decay widths as the $\psi^{\prime}$ and $J/\psi$ have the same quantum numbers except for the radial numbers   and masses, we can use the coupling constants $\delta(2,1)$ and $\delta(1,1)$
to  parameterize all those corrections in  the nonrelativistic potential models, and take them as free parameters fitted to the
 experimental data, then use those parameters to study other processes.

From the heavy quarkonium  effective Lagrangians  ${\cal L}_{SS}$, ${\cal
L}_{SP}$ and ${\cal L}_{PD}$, we can obtain the radiative decay
 widths $\Gamma$,
\begin{eqnarray}
\Gamma&=&\frac{1}{2j+1}\sum\frac{p_{cm}}{8\pi M^2 } |T|^2\, ,
\end{eqnarray}
where the $T$ denotes the scattering amplitude, the $p_{cm}$ (or
$k_\gamma$) is the momentum of the final states in the center of
mass coordinate, the $\sum$ denotes the sum of all the  polarization
vectors,  the $j$ is the total angular momentum of the initial
state, and the $M$ is the mass of the initial state. For example, in
the radiative decays $\Upsilon_3(m^3{\rm{D}_3})\to
\chi_{b2}({n}^3{\rm{P}_2})\gamma$,
\begin{eqnarray}
|T|^2&=&4M_{\Upsilon_3}M_{\chi_{2}}\delta^2(m,n)\epsilon^*_{\mu_1\nu_1\alpha_1}\epsilon^{\nu_1}_{\,\,\,\,\eta_1}\left(g^{\alpha_1\eta_1}-v^{\alpha_1}v^{\eta_1}\right)\left(k^{\mu_1}\epsilon^{\sigma_1}-k^{\sigma_1}\epsilon^{\mu_1}\right)v_{\sigma_1}
 \nonumber\\
&&\epsilon^{\mu_2\nu_2\alpha_2}\epsilon^{*\,\eta_2}_{\nu_2}\left(g_{\alpha_2\eta_2}-v_{\alpha_2}v_{\eta_2}\right)\left(k_{\mu_2}\epsilon^{*}_{\,\sigma_2}-k_{\sigma_2}\epsilon^*_{\,\mu_2}\right)v^{\sigma_2}\,,
\end{eqnarray}
where the  $\epsilon_{\mu\nu\alpha}(\lambda,q)$,
$\epsilon_{\mu\nu}(\lambda,p)$ and $\epsilon_{\mu}(\lambda,k)$ are
the polarization vectors of the bottomonium states
$\Upsilon_3({m}^3{\rm{D}_3})$,
$\chi_{b2}({n}^3{\rm{P}_2})$ and the photon, respectively.
 The summation of the  polarization vectors of the total angular momentum $j=1,2,3$
 states results in the following three formulae,
 \begin{eqnarray}
 \sum_\lambda \epsilon^*_{\mu}\epsilon_{\nu}&=&\widetilde{g}_{\mu\nu}=-g_{\mu\nu}+\frac{p_\mu p_\nu}{p^2}  \, , \nonumber\\
 \sum_\lambda \epsilon^*_{\mu\nu}\epsilon_{\alpha\beta}&=&\frac{\widetilde{g}_{\mu\alpha}\widetilde{g}_{\nu\beta}
 +\widetilde{g}_{\mu\beta}\widetilde{g}_{\nu\alpha}}{2}-\frac{\widetilde{g}_{\mu\nu}\widetilde{g}_{\alpha\beta}}{3}  \, , \nonumber\\
\sum_\lambda\epsilon^*_{\mu\nu\rho}\epsilon_{\alpha\beta\tau}&=&\frac{1}{6}\left(
\widetilde{g}_{\mu\alpha}\widetilde{g}_{\nu\beta}\widetilde{g}_{\rho\tau}
+\widetilde{g}_{\mu\alpha}\widetilde{g}_{\nu\tau}\widetilde{g}_{\rho\beta}
+\widetilde{g}_{\mu\beta}\widetilde{g}_{\nu\alpha}\widetilde{g}_{\rho\tau}
   +\widetilde{g}_{\mu\beta}\widetilde{g}_{\nu\tau}\widetilde{g}_{\rho\alpha}
   +\widetilde{g}_{\mu\tau}\widetilde{g}_{\nu\alpha}\widetilde{g}_{\rho\beta}
   +\widetilde{g}_{\mu\tau}\widetilde{g}_{\nu\beta}\widetilde{g}_{\rho\alpha}\right)\nonumber\\
   &&-\frac{1}{15}\left(\widetilde{g}_{\mu\alpha}\widetilde{g}_{\nu\rho}\widetilde{g}_{\beta\tau}
   +\widetilde{g}_{\mu\beta}\widetilde{g}_{\nu\rho}\widetilde{g}_{\alpha\tau}
   +\widetilde{g}_{\mu\tau}\widetilde{g}_{\nu\rho}\widetilde{g}_{\alpha\beta}
   +\widetilde{g}_{\nu\alpha}\widetilde{g}_{\mu\rho}\widetilde{g}_{\beta\tau}
   +\widetilde{g}_{\nu\beta}\widetilde{g}_{\mu\rho}\widetilde{g}_{\alpha\tau}
      +\widetilde{g}_{\nu\tau}\widetilde{g}_{\mu\rho}\widetilde{g}_{\alpha\beta}\right. \nonumber\\
   &&\left. +\widetilde{g}_{\rho\alpha}\widetilde{g}_{\mu\nu}\widetilde{g}_{\beta\tau}
         +\widetilde{g}_{\rho\beta}\widetilde{g}_{\mu\nu}\widetilde{g}_{\alpha\tau}
         +\widetilde{g}_{\rho\tau}\widetilde{g}_{\mu\nu}\widetilde{g}_{\alpha\beta}\right) \, ,
 \end{eqnarray}
we use the FeynCalc to carry out the contractions of the Lorentz indexes.

\section{Numerical Results}

 In calculations, the masses of the bottomonium states are taken as the
experimental values from the Belle collaboration \cite{Belle1103}, the ATLAS collaboration \cite{ATLAS2011}
and the Particle Data Group  \cite{PDG}, see Table 1; for the
unobserved bottomonium states, we take the values from the screened
potential model as the physical masses \cite{LiChao0909}. For the bottomonium states  above the $B\bar{B}$ threshold, the masses receive
contributions from the intermediate  meson-loops, such as the $B\bar{B}$,  $B\bar{B}^*$, $B^*\bar{B}$, $B^*\bar{B}^*$, $\cdots$. If the
coupled channel effects  are large enough to distort the $b\bar{b}$ configurations and induce some meson-meson components in the wave-functions,
the mass-shifts are very large. For example, in the case of the charmonium states,
the mass-shifts originate from the coupled channel effects are about $-(400-500)\,\rm{MeV}$ \cite{C-meson-loop},  we have to take the masses from the special potential quark model as the bare masses, and redefine the bare masses to reproduce the physical masses or the experimental values,
the net coupled channel effects  lead to the mass-shifts of the order $10\,\rm{MeV}$. For the observed bottomonium states, the masses from the    screened
potential model  \cite{LiChao0909}
  are consistent with the experimental data from the Belle collaboration \cite{Belle1103}, the ATLAS collaboration \cite{ATLAS2011}
and the Particle Data Group  \cite{PDG}. Although the net coupled channel effects result in mass-shifts for all the  bottomonium states, we expect the mass-shifts are not large enough to destroy counting  them  as small quantities, and  only  modify the radiative decay widths through the phase-factor $k^3_\gamma$ perturbatively. Neglecting the coupled channel effects  can lead to the  uncertainties about $\Delta k_\gamma^3/k_\gamma^3$.

 The numerical values of the radiative decay widths  are presented in Tables
 2-6, where we retain the unknown coupling constants $\delta(m,n)$
 among the multiplets of the radial quantum numbers $m$ and $n$. In general, we expect to fit
  the parameters $\delta(m,n)$  to the precise experimental data, however,
   the experimental data are far from enough in the
 present time. In Tables
 2, 7-10, we present the ratios among the radiative transitions  among the
 bottomonium states.

 The CLEO collaboration have observed that the doubly  radiative decay $\psi^{\prime}\to  \gamma\gamma{J/\psi}$ takes place through
 the decay cascade $\psi^{\prime}\to \gamma\chi_{cj} $, $\chi_{cj}\to \gamma {J/\psi} $, $j=0,1,2$ with additional  tiny non-resonance contributions \cite{CLEO-05,CLEO-08}.  Recently, the BESIII collaboration observed  the first evidence for the direct two-photon transition $\psi^{\prime}\to \gamma\gamma J/\psi $ with the branching fraction     $(3.3\pm0.6 {}^{+0.8}_{-1.1} ) \times10^{-4}$ in a sample of 106 million $\psi^{\prime}$ decays collected by the BESIII detector \cite{BES1204}. In Ref.\cite{He1012}, He et al study the discrete contributions to decay $\psi^{\prime}\to \gamma\gamma J/\psi$  due
to the $E_1$ transitions using the heavy quarkonium effective Lagrangian \cite{PRT1997}.
 We expect that the corresponding doubly  radiative decay $\Upsilon(2{\rm S})\to  \gamma\gamma{\Upsilon(1{\rm S})}$ occurs through the analogous decay cascade
  $\Upsilon(2{\rm S})\to \gamma\chi_{bj}(1{\rm P}) $, $\chi_{bj}(1{\rm P})\to \gamma \Upsilon(1{\rm S}) $. Experimentally, the doubly radiative decays $\Upsilon(3{\rm S}) \to \gamma\gamma \Upsilon(2{\rm S})$ and $\Upsilon_2(1{\rm D}) \to \gamma\gamma \Upsilon(1{\rm S})$ have been observed \cite{PDG}. Once the coupling constants $\delta(m,n)$ in the heavy quarkonium effective  Lagrangians ${\cal{L}}_{SS}$, ${\cal{L}}_{SP}$ and
${\cal{L}}_{PD}$ are fitted to the precise  experimental data, we can use them to study the doubly radiative decays,  $\Upsilon(3{\rm S}) \to \gamma\gamma \Upsilon(2{\rm S})$, $\Upsilon_2(1{\rm D}) \to \gamma\gamma \Upsilon(1{\rm S})$, and other physical processes have  the radiative transitions  as their  sub-processes, or study the singly radiative decays. For example, the radiative decays $\Upsilon({\rm 2S}) \to \gamma \chi_{bj}({\rm 1P})$ and $\Upsilon({\rm 3S}) \to \gamma \chi_{bj}({\rm 2P})$ have been observed experimentally,  we can use them to fit the coupling constants $\delta(2,1)$ and $\delta(3,2)$, and make predictions for the decay widths $\Gamma(\eta_b({\rm 2S}) \to \gamma h_{b}({\rm 1P}))$ and $\Gamma(\eta_b({\rm 3S}) \to \gamma h_{b}({\rm 2P}))$, see Tables 3,7.

 The  widths of the radiative transitions of the $S$-wave  to the $P$-wave bottomonium
 states listed in the Review of Particle Physics are presented in Table 3 \cite{PDG}. From those radiative decay widths,
 we can obtain the ratios among the radiative decay widths of the $S$-wave
 to the $P$-wave bottomonium states,  which are
presented in Table 7.  From the table, we can see that the
agreements between the experimental data and the theoretical
predictions are rather good, and the heavy quarkonium  effective theory in
the leading order approximation  works rather well. The ratios
presented in Tables 2, 7-10 can be confronted with the experimental
data in the future at the Tevatron, KEK-B, RHIC
 and LHCb. In Tables 7-10, we also present values come from  the    screened
potential model for comparison   \cite{LiChao0909}, and no definite conclusion can be made.

  In calculations, we observe that the radiative decay
widths are sensitive  to the masses of the bottomonium states in some channels. The
experimental value of the $\Upsilon(4 {\rm{S}})$ from the Particle
Data Group  is $M_{\Upsilon(4 {\rm{S}})}=(10579.4 \pm
1.2)\,\rm{MeV}$ \cite{PDG}, while the predictions of  the $4{\rm S}$
states from the screened potential model are $M_{\Upsilon(4
{\rm{S}})}=10661\,\rm{MeV}$ and $M_{\eta_b(4
{\rm{S}})}=10595\,\rm{MeV}$ \cite{LiChao0909}. In this article, we
take the values $M_{\Upsilon(4 {\rm{S}})}=(10579.4 \pm
1.2)\,\rm{MeV}$ and $M_{\eta_b(4 {\rm{S}})}=10595\,\rm{MeV}$, i.e.
$M_{\Upsilon(4 {\rm{S}})}<M_{\eta_b(4 {\rm{S}})}$, the prediction of
the ratio $\frac{\Gamma(\eta_b(4{\rm S}) \to h_b(3 {\rm P})
\gamma)}{\Gamma(\Upsilon(4{\rm S})\to\chi_{b2}(3 {\rm
P})\gamma)}=6.297$ seems rather exotic. Naively, we expect that  the
spin-1 states $\Upsilon({\rm nS})$ have larger masses than the
corresponding ones of the spin-0 states $\eta_b({\rm nS})$. If we
take $M_{\Upsilon(4 {\rm{S}})}>M_{\eta_b(4 {\rm{S}})}$, then the
ratio $\frac{\Gamma(\eta_b(4{\rm S}) \to h_b(3 {\rm P})
\gamma)}{\Gamma(\Upsilon(4{\rm S})\to\chi_{b2}(3 {\rm
P})\gamma)}\sim 3$ seems rather natural, more experimental data are
still needed to make better predictions.

The radiative decay widths $\Gamma \propto k_\gamma^3$,  the
uncertainties originate from the masses of the bottomonium states
can be estimated as
\begin{eqnarray}
\frac{\Delta\Gamma}{\Gamma} &\approx& \frac{\Delta
k_\gamma^3}{k_\gamma^3} =\frac{3}{1-\frac{M_f^2}{M_i^2}}\sqrt{
\left(1+\frac{M_f^2}{M_i^2}\right)^2\left(\frac{\Delta
M_i}{M_i}\right)^2+4\left(\frac{M_f^2}{M_i^2}\right)^2\left(\frac{\Delta
M_f}{M_f}\right)^2}\, , \nonumber \\
&<&\frac{6}{1-\frac{M_f^2}{M_i^2}}\sqrt{ \left(\frac{\Delta
M_i}{M_i}\right)^2+\left(\frac{\Delta M_f}{M_f}\right)^2}\, ,
\end{eqnarray}
where we have neglected the terms $\frac{\Delta M_i}{M_i}$ and
$\frac{\Delta M_f}{M_f}$  not enhanced by the factor
$\frac{1}{1-{M_f^2}/{M_i^2}}$, the  subscripts  $i$ and $f$ denote
the initial and final bottomonium states, respectively. The
uncertainties of the masses of the bottomonium states
$\Upsilon(5^3{\rm{S}}_1)$ and $\Upsilon(6^3{\rm{S}}_1)$  are
$8\,\rm{MeV}$ from the  Particle Data Group \cite{PDG}, and the
corresponding relative uncertainties $\frac{\Delta M }{M}$ are
$0.074\%$ and $0.073\%$, respectively, and expected to result in the
largest uncertainties for the decay widths. The  ATLAS collaboration fixed  the hyperfine mass splitting between the $\chi_{b2}(\rm{3P})$ and $\chi_{b1}(\rm{3P})$
to the theoretically predicted value of $12\,\rm{ MeV}$, which maybe result in larger uncertainty  for the mass barycenter,  as
 the theoretical and experimental values of the  hyperfine mass splitting always have differences \cite{ATLAS2011}.
On the other hand, the
uncertainties of the masses of other bottomonium states are very
small, about (or less than) $1\,\rm{MeV}$, from the recent Belle
data \cite{Belle1103} and the Review of Particle Physics \cite{PDG},
the relative uncertainties $\frac{\Delta M }{M}$ are tiny. If the
factor $\frac{1}{1-{M_f^2}/{M_i^2}}$ is not large enough (i.e. the
difference between the radial quantum numbers of the initial ($m$)
and final ($n$) states is larger than 1, $|m-n|>1$), the
uncertainties originate from the masses of the bottomonium states
are of a few percents, and can be neglected. For example, the uncertainties  in the
radiative decays of the $S$-wave to the $S$-wave and the $S$-wave to the $P$-wave bottomonium
states are
\begin{eqnarray}
  \frac{\Delta\Gamma}{\Gamma}<0.9\% &{\rm{for}}&\Upsilon(4^3{\rm{S}}_1)\to\eta_b(1^1{\rm{S}}_0)\,\gamma\,,\nonumber\\
  \frac{\Delta\Gamma}{\Gamma}<0.6\% &{\rm{for}}&\Upsilon(4^3{\rm{S}}_1)\to \eta_b(2^1{\rm{S}}_0)\,\gamma\, ,\nonumber\\
  \frac{\Delta\Gamma}{\Gamma}<1.5\% &{\rm{for}}&\Upsilon(4^3{\rm{S}}_1)\to\eta_b(3^1{\rm{S}}_0)\,\gamma\, ,\nonumber\\
  \frac{\Delta\Gamma}{\Gamma}<1.9\% &{\rm{for}}&\Upsilon(5^3{\rm{S}}_1)\to\eta_b(1^1{\rm{S}}_0)\,\gamma\, ,\nonumber\\
  \frac{\Delta\Gamma}{\Gamma}<2.9\% &{\rm{for}}&\Upsilon(5^3{\rm{S}}_1)\to\eta_b(2^1{\rm{S}}_0)\,\gamma\, ,\nonumber\\
  \frac{\Delta\Gamma}{\Gamma}<4.6\% &{\rm{for}}&\Upsilon(5^3{\rm{S}}_1)\to\eta_b(3^1{\rm{S}}_0)\,\gamma\, ,\nonumber\\
  \frac{\Delta\Gamma}{\Gamma}<9.0\% &{\rm{for}}&\Upsilon(5^3{\rm{S}}_1)\to\eta_b(4^1{\rm{S}}_0)\,\gamma\,  ,
\end{eqnarray}
and
\begin{eqnarray}
  \frac{\Delta\Gamma}{\Gamma}<0.6\% &{\rm{for}}&\Upsilon(4^3{\rm{S}}_1)\to\chi_{b1}(1^3{\rm{P}}_1)\,\gamma\,,\nonumber\\
  \frac{\Delta\Gamma}{\Gamma}<1.3\% &{\rm{for}}&\Upsilon(4^3{\rm{S}}_1)\to \chi_{b1}(2^3{\rm{P}}_1)\,\gamma\, ,\nonumber\\
  \frac{\Delta\Gamma}{\Gamma}<6.5\% &{\rm{for}}&\Upsilon(4^3{\rm{S}}_1)\to\chi_{b1}(3^3{\rm{P}}_1)\,\gamma\, ,\nonumber\\
  \frac{\Delta\Gamma}{\Gamma}<2.6\% &{\rm{for}}&\Upsilon(5^3{\rm{S}}_1)\to\chi_{b1}(1^3{\rm{P}}_1)\,\gamma\, ,\nonumber\\
  \frac{\Delta\Gamma}{\Gamma}<3.9\% &{\rm{for}}&\Upsilon(5^3{\rm{S}}_1)\to\chi_{b1}(2^3{\rm{P}}_1)\,\gamma\, ,\nonumber\\
  \frac{\Delta\Gamma}{\Gamma}<7.2\% &{\rm{for}}&\Upsilon(5^3{\rm{S}}_1)\to\chi_{b1}(3^3{\rm{P}}_1)\,\gamma\, ,\nonumber\\
  \frac{\Delta\Gamma}{\Gamma}<22\% &{\rm{for}}&\Upsilon(5^3{\rm{S}}_1)\to\chi_{b1}(4^3{\rm{P}}_1)\,\gamma\,  ,
\end{eqnarray}
respectively. From above equations, we can see that the relative uncertainties of
the decay widths of the $\Upsilon(5^3{\rm{S}}_1) \to
\eta_b({n}^1{\rm{S}}_0), \, \chi_{b1}({n}^3{\rm{P}}_1)$ are larger than the corresponding ones
of the $\Upsilon(4^3{\rm{S}}_1)\to \eta_b({n}^1{\rm{S}}_0), \, \chi_{b1}({n}^3{\rm{P}}_1)$, as
the mass of the $\Upsilon(5^3{\rm{S}}_1)$ has larger uncertainty
\cite{PDG}. In calculations, we observe that the upper bound of the
uncertainties originate from the masses of the
$\Upsilon(4^3{\rm{S}}_1)$ and $\Upsilon(5^3{\rm{S}}_1)$ are about
ten (or twenty) percent for the values $|m-n|\leq 1$.

 In this article,
we have neglected the corrections of the order $\mathcal{O}(1/m^2_Q)$ and $\mathcal{O}(1/m_Q)$ for the heavy quarkonium effective Lagrangians
$\mathcal{L}_{SS}$ (as the spin-flipped Lagrangian $\mathcal{L}_{SS}$ is of the
order $\mathcal{O}(1/m_Q)$) and $\mathcal{L}_{SP}$, $\mathcal{L}_{PD}$, respectively,
which maybe result in rather large  uncertainties. If those corrections can be counted as $\Lambda_{QCD}/m_b$,  taking
 the  Lagrangians $\mathcal{L}_{SS}$, $\mathcal{L}_{SP}$ and $\mathcal{L}_{PD}$   to fit the experimental data to determine the unknown coupling
constants can result in uncertainties of the order   $\mathcal{O}(\Lambda_{QCD}/m_b)$. Such crude estimation maybe not work well,
we should bear in mind that the magnitudes of the contributions from the higher order terms in the heavy quarkonium  effective theory
 be determined experimentally.

\begin{table}
\begin{center}
\begin{tabular}{|c|c|c|c|c| }\hline\hline
    & $\Gamma(\Upsilon \to\eta_b \gamma)$& $\Gamma(\eta_b\to \Upsilon\gamma)$ & $\frac{\Gamma(\eta_b\to \Upsilon\gamma)}{\Gamma(\Upsilon \to\eta_b \gamma)}$   \\ \hline
                   $2{\rm S}\to1{\rm S}$       & 0.091           & 0.163          &  1.781  \\ \hline

                   $3{\rm S}\to1{\rm S}$       & 0.299           & 0.674          &  2.254  \\ \hline
                   $3{\rm S}\to2{\rm S}$       & 0.019           & 0.034          &  1.761  \\ \hline

                   $4{\rm S}\to1{\rm S}$       & 0.532           & 1.408          &  2.648  \\ \hline
                   $4{\rm S}\to2{\rm S}$       & 0.076           & 0.207          &  2.711  \\ \hline
                   $4{\rm S}\to3{\rm S}$       & 0.006           & 0.017          &  2.673  \\ \hline

                   $5{\rm S}\to1{\rm S}$       & 0.952           & 2.290          &  2.406  \\ \hline
                   $5{\rm S}\to2{\rm S}$       & 0.233           & 0.527          &  2.261  \\ \hline
                   $5{\rm S}\to3{\rm S}$       & 0.057           & 0.113          &  1.962  \\ \hline
                   $5{\rm S}\to4{\rm S}$       & 0.008           & 0.016          &  2.059  \\ \hline

                   $6{\rm S}\to1{\rm S}$       & 1.240           & 3.277          &  2.643  \\ \hline
                   $6{\rm S}\to2{\rm S}$       & 0.366           & 0.973          &  2.658  \\ \hline
                   $6{\rm S}\to3{\rm S}$       & 0.118           & 0.308          &  2.607  \\ \hline
                   $6{\rm S}\to4{\rm S}$       & 0.029           & 0.093          &  3.158  \\ \hline
                   $6{\rm S}\to5{\rm S}$       & 0.003           & 0.004          &  1.147  \\ \hline

                   $7{\rm S}\to1{\rm S}$       & 1.620           & 4.330          &  2.673  \\ \hline
                   $7{\rm S}\to2{\rm S}$       & 0.563           & 1.517          &  2.697  \\ \hline
                   $7{\rm S}\to3{\rm S}$       & 0.224           & 0.597          &  2.669  \\ \hline
                   $7{\rm S}\to4{\rm S}$       & 0.079           & 0.244          &  3.081 \\ \hline
                   $7{\rm S}\to5{\rm S}$       & 0.021           & 0.038          &  1.839 \\ \hline
                   $7{\rm S}\to6{\rm S}$       & 0.002           & 0.005          &  2.204 \\ \hline  \hline
\end{tabular}
\end{center}
\caption{ The  ratios of the radiative decay widths of the $S$-wave
 to the $S$-wave bottomonium states, where the unit of the widths is
$\delta^2(m,n)$. }
\end{table}

\begin{table}
\begin{center}
\begin{tabular}{|c|c|c|c|c|c| }\hline\hline
             $\Gamma$                    & $\Upsilon \to \chi_{2}\gamma$& $\Upsilon \to \chi_{1}\gamma$ & $\Upsilon \to \chi_{0}\gamma$ & $\eta_b \to h_b\gamma$  \\ \hline
           $2{\rm S}\to1{\rm P}$         & 2.355                   & 2.281                     & 1.492                     & 2.176  \\ \hline
 $\widehat{2{\rm S}\to1{\rm P}}$         & $2.287\pm0.219$         & $2.207\pm0.222$           & $1.215\pm0.162$           &  \\ \hline

           $3{\rm S}\to1{\rm P}$         & 138.0                   & 93.76                     & 38.16                     & 230.4  \\ \hline
 $\widehat{3{\rm S}\to1{\rm P}}$         & $<0.386$                & $<0.035$                  & 0.061                     &  \\ \hline
           $3{\rm S}\to2{\rm P}$         & 1.123                   & 1.028                     & 0.634                     & 1.084  \\ \hline
 $\widehat{3{\rm S}\to2{\rm P}}$         & $2.662\pm0.406$         & $2.560\pm0.337$           & $1.199\pm0.164$           &    \\ \hline

           $4{\rm S}\to1{\rm P}$         & 447.6                   & 291.6                     & 110.9                     & 909.9   \\ \hline
           $4{\rm S}\to2{\rm P}$         & 49.28                   & 33.40                     & 13.59                     & 110.7   \\ \hline
           $4{\rm S}\to3{\rm P}$         & 0.143                   & 0.178                     & 0.161                     & 0.901  \\ \hline

           $5{\rm S}\to1{\rm P}$         & 1223                    & 777.3                     & 283.1                     & 1983    \\ \hline
           $5{\rm S}\to2{\rm P}$         & 326.4                   & 208.6                     & 77.14                     & 483.1  \\ \hline
           $5{\rm S}\to3{\rm P}$         & 58.35                   & 38.88                     & 15.55                     & 71.10   \\ \hline
           $5{\rm S}\to4{\rm P}$         & 1.627                   & 1.453                     & 0.807                     & 0.678   \\ \hline

           $6{\rm S}\to1{\rm P}$         & 1854                    & 1169                      & 419.5                     & 3375  \\ \hline
           $6{\rm S}\to2{\rm P}$         & 628.4                   & 396.5                     & 143.3                     & 1133    \\ \hline
           $6{\rm S}\to3{\rm P}$         & 178.4                   & 114.9                     & 43.38                     & 318.9   \\ \hline
           $6{\rm S}\to4{\rm P}$         & 26.72                   & 18.80                     & 7.829                     & 49.22   \\ \hline
           $6{\rm S}\to5{\rm P}$         & 0.275                   & 0.329                     & 0.221                     & 0.552   \\ \hline

           $7{\rm S}\to1{\rm P}$         & 2770                    & 1736                      & 614.5                     & 5003   \\ \hline
           $7{\rm S}\to2{\rm P}$         & 1132                    & 707.3                     & 251.2                     & 2025    \\ \hline
           $7{\rm S}\to3{\rm P}$         & 432.2                   & 273.2                     & 99.66                     & 766.9  \\ \hline
           $7{\rm S}\to4{\rm P}$         & 124.2                   & 81.88                     & 31.21                     & 223.4   \\ \hline
           $7{\rm S}\to5{\rm P}$         & 19.91                   & 14.24                     & 5.863                     & 35.84   \\ \hline \hline

\end{tabular}
\end{center}
\caption{ The radiative decay widths of the $S$-wave   to the
$P$-wave bottomonium states, where the unit is
$10^{-4}\delta^2(m,n)$. The wide-hat denotes the experimental
values, where the unit is KeV. }
\end{table}

\begin{table}
\begin{center}
\begin{tabular}{|c|c|c|c|c|c| }\hline\hline
                  $\Gamma$             & $\chi_{2} \to \Upsilon \gamma$& $ \chi_{1}\to \Upsilon\gamma$ & $\chi_{0} \to \Upsilon\gamma$ & $h_b \to \eta_b\gamma$  \\ \hline
            $1{\rm P}\to1{\rm S}$      & 87.27          & 76.89          & 60.89          & 121.6    \\ \hline

            $2{\rm P}\to1{\rm S}$      & 458.6          & 438.5          & 402.5          & 559.5    \\ \hline
            $2{\rm P}\to2{\rm S}$      & 14.76          & 12.55          & 9.232          & 20.13     \\ \hline

            $3{\rm P}\to1{\rm S}$      & 1017           & 988.1          & 927.5          & 1181    \\ \hline
            $3{\rm P}\to2{\rm S}$      & 126.5          & 118.2          & 103.7          & 148.2    \\ \hline
            $3{\rm P}\to3{\rm S}$      & 6.006          & 4.902          & 3.241          & 7.973     \\ \hline

            $4{\rm P}\to1{\rm S}$      & 1733           & 1688           & 1602           & 1940    \\ \hline
            $4{\rm P}\to2{\rm S}$      & 366.3          & 347.4          & 319.0          & 403.1     \\ \hline
            $4{\rm P}\to3{\rm S}$      & 67.28          & 60.86          & 51.94          & 74.70    \\ \hline
            $4{\rm P}\to4{\rm S}$      & 6.706          & 5.331          & 3.638          & 4.343     \\ \hline

            $5{\rm P}\to1{\rm S}$      & 2536           & 2484           & 2379           & 2787    \\ \hline
            $5{\rm P}\to2{\rm S}$      & 711.8          & 683.7          & 644.7          & 766.1     \\ \hline
            $5{\rm P}\to3{\rm S}$      & 209.0          & 195.7          & 178.9          & 224.0    \\ \hline
            $5{\rm P}\to4{\rm S}$      & 55.67          & 50.00          & 43.23          & 45.56    \\ \hline
            $5{\rm P}\to5{\rm S}$      & 1.037          & 0.662          & 0.328          & 2.702    \\ \hline   \hline
\end{tabular}
\end{center}
\caption{ The radiative decay widths of the $P$-wave   to the
$S$-wave bottomonium states, where the unit is
$10^{-4}\delta^2(m,n)$. }
\end{table}

\begin{table}
\begin{center}
\begin{tabular}{|c|c|c|c|c|c|c|c|c| }\hline\hline
            $\Gamma$         & $\chi_{2} \to\Upsilon_3 \gamma$& $ \chi_{2}\to \Upsilon_2\gamma$ & $\chi_{2} \to \Upsilon\gamma$ & $\chi_{1} \to \Upsilon_{2}\gamma$ & $\chi_{1} \to \Upsilon\gamma$ & $\chi_{0} \to \Upsilon\gamma$ & $h_b \to \eta_2\gamma$ \\ \hline
    $2{\rm P}\to1{\rm D}$    & 2.066        & 0.321        & 0.032        & 1.089       & 0.580       & 1.160      &   2.155 \\ \hline

    $3{\rm P}\to1{\rm D}$    & 74.46        & 12.79        & 0.963        & 58.12       & 21.98       & 73.83      &   86.58\\ \hline
    $3{\rm P}\to2{\rm D}$    & 1.207        & 0.244        & 0.019        & 0.827       & 0.337       & 0.597      &   1.262\\ \hline

    $4{\rm P}\to1{\rm D}$    & 293.8        & 51.27        & 3.676        & 239.4       & 86.16       & 311.1      &   339.6  \\ \hline
    $4{\rm P}\to2{\rm D}$    & 47.27        & 8.752        & 0.615        & 38.51       & 13.57       & 44.49      &   52.79 \\ \hline
    $4{\rm P}\to3{\rm D}$    & 0.959        & 0.196        & 0.016        & 0.595       & 0.248       & 0.415      &   0.889 \\ \hline

    $5{\rm P}\to1{\rm D}$    & 653.1        & 114.8        & 8.061        & 544.6       & 192.2       & 719.2      &   759.9  \\ \hline
    $5{\rm P}\to2{\rm D}$    & 188.5        & 34.43        & 2.368        & 159.1       & 54.85       & 197.9      &   215.7 \\ \hline
    $5{\rm P}\to3{\rm D}$    & 32.22        & 5.995        & 0.424        & 25.90       & 9.198       & 30.28      &   35.64  \\ \hline
    $5{\rm P}\to4{\rm D}$    & 0.719        & 0.149        & 0.012        & 0.429       & 0.176       & 0.307      &   0.649  \\ \hline \hline
\end{tabular}
\end{center}
\caption{ The radiative decay widths of the $P$-wave   to the
$D$-wave bottomonium states, where the unit is
$10^{-4}\delta^2(m,n)$.  }
\end{table}

\begin{table}
\begin{center}
\begin{tabular}{|c|c|c|c|c|c|c|c|c| }\hline\hline
    $\Gamma$                 & $ \Upsilon_3  \to \chi_{2}\gamma$& $ \Upsilon_2\to  \chi_{2}\gamma$ & $\Upsilon_{2} \to  \chi_{1}\gamma$ & $\Upsilon \to \chi_{2}\gamma$ & $\Upsilon \to \chi_{1}\gamma$ & $ \Upsilon\to \chi_{0}\gamma$ & $\eta_2 \to h_b\gamma$ \\ \hline
    $1{\rm D}\to1{\rm P}$    & 14.48        & 3.847       & 14.38      & 0.351        & 6.664       & 12.78      &   16.28   \\ \hline

    $2{\rm D}\to1{\rm P}$    & 138.9        & 34.00       & 113.0      & 3.638        & 60.85       & 96.19      &   147.2  \\ \hline
    $2{\rm D}\to2{\rm P}$    & 5.302        & 1.237       & 4.632      & 0.123        & 2.331       & 4.460      &   5.852   \\ \hline

    $3{\rm D}\to1{\rm P}$    & 400.7        & 98.87       & 317.3      & 10.64        & 172.7       & 258.3      &   416.3   \\ \hline
    $3{\rm D}\to2{\rm P}$    & 67.05        & 16.30       & 53.59      & 1.730        & 28.56       & 44.49      &   69.83   \\ \hline
    $3{\rm D}\to3{\rm P}$    & 3.063        & 0.704       & 2.697      & 0.069        & 1.330       & 2.687      &   3.322   \\ \hline

    $4{\rm D}\to1{\rm P}$    & 781.8        & 193.7       & 610.7      & 20.89        & 335.2       & 487.6      &   804.2  \\ \hline
    $4{\rm D}\to2{\rm P}$    & 216.5        & 53.17       & 169.2      & 5.737        & 91.91       & 135.6      &   222.2  \\ \hline
    $4{\rm D}\to3{\rm P}$    & 41.97        & 10.15       & 33.61      & 1.079        & 17.92       & 28.49      &   43.37   \\ \hline
    $4{\rm D}\to4{\rm P}$    & 1.740        & 0.393       & 1.658      & 0.038        & 0.819       & 1.738      &   2.062   \\ \hline

    $5{\rm D}\to1{\rm P}$    & 1261         & 313.5       & 977.0      & 33.69        & 537.8       & 769.1      &   1289     \\ \hline
    $5{\rm D}\to2{\rm P}$    & 453.4        & 111.9       & 350.5      & 12.10        & 191.6       & 275.6      &   462.0    \\ \hline
    $5{\rm D}\to3{\rm P}$    & 142.4        & 34.85       & 111.4      & 3.751        & 60.26       & 90.07      &   145.3    \\ \hline
    $5{\rm D}\to4{\rm P}$    & 27.29        & 6.563       & 22.51      & 0.693        & 11.93       & 19.30      &   29.18    \\ \hline
    $5{\rm D}\to5{\rm P}$    & 1.166        & 0.259       & 1.149      & 0.025        & 0.559       & 1.173      &   1.415   \\ \hline \hline
\end{tabular}
\end{center}
\caption{ The radiative decay widths of the $D$-wave   to the
$P$-wave bottomonium states, where the unit is
$10^{-4}\delta^2(m,n)$.  }
\end{table}

\begin{table}
\begin{center}
\begin{tabular}{|c|c|c|c|c| }\hline\hline
            $\widetilde{\Gamma}$    &$\Upsilon \to \chi_{1}\gamma$ & $\Upsilon \to  \chi_{0}\gamma$ & $\eta_b \to h_b\gamma$  \\ \hline
           $2{\rm S}\to1{\rm P}$           & 0.969 [0.846]     & 0.634 [0.451]     & 0.924 [2.264] \\ \hline
 $\widehat{2{\rm S}\to1{\rm P}}$           & $0.965\pm0.134$   & $0.531\pm0.087$   &       \\ \hline

           $3{\rm S}\to1{\rm P}$           & 0.679 [0.111]     & 0.277 [0.040]     & 1.670 [4.508] \\ \hline
           $3{\rm S}\to2{\rm P}$           & 0.915 [0.803]     & 0.565 [0.405]     & 0.966 [3.322]  \\ \hline
 $\widehat{3{\rm S}\to2{\rm P}}$           & $0.962\pm0.194$   & $0.450\pm0.092$   &    \\ \hline

           $4{\rm S}\to1{\rm P}$           & 0.651 [0.233]     & 0.248 [0.001]     & 2.033 [6.558]  \\ \hline
           $4{\rm S}\to2{\rm P}$           & 0.678 [0.002]     & 0.276 [0.375]     & 2.246 [3.857] \\ \hline
           $4{\rm S}\to3{\rm P}$           & 1.245 [1.423]     & 1.125 [1.038]     & 6.297 [24.81] \\ \hline

           $5{\rm S}\to1{\rm P}$           & 0.636             & 0.231             & 1.622  \\ \hline
           $5{\rm S}\to2{\rm P}$           & 0.639             & 0.236             & 1.480  \\ \hline
           $5{\rm S}\to3{\rm P}$           & 0.666             & 0.266             & 1.218  \\ \hline
           $5{\rm S}\to4{\rm P}$           & 0.893             & 0.496             & 0.417  \\ \hline

           $6{\rm S}\to1{\rm P}$           & 0.631             & 0.226             & 1.820  \\ \hline
           $6{\rm S}\to2{\rm P}$           & 0.631             & 0.228             & 1.803  \\ \hline
           $6{\rm S}\to3{\rm P}$           & 0.644             & 0.243             & 1.788  \\ \hline
           $6{\rm S}\to4{\rm P}$           & 0.703             & 0.293             & 1.842   \\ \hline
           $6{\rm S}\to5{\rm P}$           & 1.194             & 0.802             & 2.007  \\ \hline

           $7{\rm S}\to1{\rm P}$           & 0.627             & 0.222             & 1.806  \\ \hline
           $7{\rm S}\to2{\rm P}$           & 0.625             & 0.222             & 1.790  \\ \hline
           $7{\rm S}\to3{\rm P}$           & 0.632             & 0.231             & 1.775  \\ \hline
           $7{\rm S}\to4{\rm P}$           & 0.659             & 0.251             & 1.799  \\ \hline
           $7{\rm S}\to5{\rm P}$           & 0.715             & 0.294             & 1.800  \\ \hline \hline

\end{tabular}
\end{center}
\caption{ The ratios among the radiative decay widths of the
$S$-wave   to the $P$-wave bottomonium states, where
$\widetilde{\Gamma}=\frac{\Gamma}{ \Gamma(\Upsilon\to
\chi_{2}\gamma)}$, $\widetilde{\Gamma}(\Upsilon\to\chi_{2}\gamma)=\frac{\Gamma(\Upsilon\to\chi_{2}\gamma)}{ \Gamma(\Upsilon\to\chi_{2}\gamma)}=1$,
 and the wide-hat denotes the experimental values, the values in the bracket come from the screened potential model \cite{LiChao0909}.}
\end{table}

\begin{table}
\begin{center}
\begin{tabular}{|c|c|c|c|c| }\hline\hline
        $\widetilde{\Gamma}$       & $ \chi_{1}\to \Upsilon\gamma$ & $\chi_{0} \to \Upsilon\gamma$ & $h_b \to \eta_b\gamma$  \\ \hline
        $1{\rm P}\to1{\rm S}$       & 0.881 [0.920]  & 0.698 [0.745]  & 1.394 [1.114]    \\ \hline

        $2{\rm P}\to1{\rm S}$       & 0.956 [0.685]  & 0.878 [0.360]  & 1.220 [1.440]     \\ \hline
        $2{\rm P}\to2{\rm S}$       & 0.850 [0.972]  & 0.625 [0.817]  & 1.364 [1.077]  \\ \hline

        $3{\rm P}\to1{\rm S}$       & 0.972 [0.501]  & 0.912 [0.127]  & 1.161 [1.399]     \\ \hline
        $3{\rm P}\to2{\rm S}$       & 0.934 [0.782]  & 0.820 [0.533]  & 1.172 [1.495]  \\ \hline
        $3{\rm P}\to3{\rm S}$       & 0.816 [0.898]  & 0.540 [0.691]  & 1.327 [1.045]    \\ \hline

        $4{\rm P}\to1{\rm S}$       & 0.974          & 0.924          & 1.120    \\ \hline
        $4{\rm P}\to2{\rm S}$       & 0.948          & 0.871          & 1.101    \\ \hline
        $4{\rm P}\to3{\rm S}$       & 0.905          & 0.772          & 1.110    \\ \hline
        $4{\rm P}\to4{\rm S}$       & 0.795          & 0.543          & 0.648   \\ \hline

        $5{\rm P}\to1{\rm S}$       & 0.979          & 0.938          & 1.099    \\ \hline
        $5{\rm P}\to2{\rm S}$       & 0.960          & 0.906          & 1.076     \\ \hline
        $5{\rm P}\to3{\rm S}$       & 0.936          & 0.856          & 1.072    \\ \hline
        $5{\rm P}\to4{\rm S}$       & 0.898          & 0.776          & 0.818   \\ \hline
        $5{\rm P}\to5{\rm S}$       & 0.638          & 0.317          & 2.605    \\ \hline   \hline
\end{tabular}
\end{center}
\caption{ The ratios among the radiative decay widths of the
$P$-wave   to the $S$-wave bottomonium states, where
$\widetilde{\Gamma}=\frac{\Gamma}{\Gamma(\chi_{2} \to
\Upsilon\gamma)}$, $\widetilde{\Gamma}(\chi_{2} \to\Upsilon\gamma)=\frac{\Gamma(\chi_{2} \to\Upsilon\gamma)}{\Gamma(\chi_{2} \to\Upsilon\gamma)}=1$,
and the values in the bracket come from the screened potential model \cite{LiChao0909} }
\end{table}

\begin{table}
\begin{center}
\begin{tabular}{|c|c|c|c|c|c|c|c| }\hline\hline
    $\widetilde{\Gamma}$     & $ \chi_{2}\to \Upsilon_2\gamma$ & $\chi_{2} \to \Upsilon\gamma$ & $\chi_{1} \to \Upsilon_{2}\gamma$ & $\chi_{1} \to \Upsilon\gamma$ & $\chi_{0} \to \Upsilon\gamma$ & $h_b \to \eta_2\gamma$ \\ \hline
    $2{\rm P}\to1{\rm D}$      & 0.156 [0.185] & 0.016 [0.013] & 0.527 [0.722] & 0.281 [0.268] & 0.561 [0.591] & 1.043 [2.371] \\ \hline

    $3{\rm P}\to1{\rm D}$      & 0.172         & 0.013         & 0.781         & 0.295         & 0.992         & 1.163\\ \hline
    $3{\rm P}\to2{\rm D}$      & 0.202 [0.190] & 0.016 [0.015] & 0.685 [0.635] & 0.279 [0.241] & 0.494 [0.471] & 1.046 [2.388]\\ \hline

    $4{\rm P}\to1{\rm D}$      & 0.175        & 0.013        & 0.815       & 0.293       & 1.059      &   1.156  \\ \hline
    $4{\rm P}\to2{\rm D}$      & 0.185        & 0.013        & 0.815       & 0.287       & 0.941      &   1.117  \\ \hline
    $4{\rm P}\to3{\rm D}$      & 0.204        & 0.016        & 0.620       & 0.259       & 0.433      &   0.927  \\ \hline

    $5{\rm P}\to1{\rm D}$      & 0.176        & 0.012        & 0.834       & 0.294       & 1.101      &   1.164  \\ \hline
    $5{\rm P}\to2{\rm D}$      & 0.183        & 0.013        & 0.844       & 0.291       & 1.050      &   1.144  \\ \hline
    $5{\rm P}\to3{\rm D}$      & 0.186        & 0.013        & 0.804       & 0.285       & 0.940      &   1.106  \\ \hline
    $5{\rm P}\to4{\rm D}$      & 0.207        & 0.016        & 0.596       & 0.245       & 0.426      &   0.903 \\ \hline \hline
\end{tabular}
\end{center}
\caption{ The ratios among the radiative decay widths of the
$P$-wave   to the $D$-wave bottomonium states, where
$\widetilde{\Gamma}=\frac{\Gamma}{\Gamma(\chi_{2} \to\Upsilon_3
\gamma)}$,  $\widetilde{\Gamma}(\chi_{2} \to\Upsilon_3\gamma)=\frac{\Gamma(\chi_{2} \to\Upsilon_3\gamma)}{\Gamma(\chi_{2} \to\Upsilon_3\gamma)}=1$,
and the values in the bracket come from the screened potential model \cite{LiChao0909} }
\end{table}

\begin{table}
\begin{center}
\begin{tabular}{|c|c|c|c|c|c|c|c| }\hline\hline
    $\widetilde{\Gamma}$     & $ \Upsilon_2\to  \chi_{2}\gamma$ & $\Upsilon_{2} \to  \chi_{1}\gamma$ & $\Upsilon \to \chi_{2}\gamma$ & $\Upsilon \to \chi_{1}\gamma$ & $ \Upsilon\to \chi_{0}\gamma$ & $\eta_2 \to h_b\gamma$ \\ \hline
    $1{\rm D}\to1{\rm P}$      & 0.266 [0.240] & 0.994 [0.808] & 0.024 [0.025] & 0.460 [0.420] & 0.883 [0.682] & 1.125 [1.490]  \\ \hline

    $2{\rm D}\to1{\rm P}$      & 0.245 [0.182] & 0.814 [1.196] & 0.026 [0.013] & 0.438 [0.501] & 0.693 [1.662] & 1.060 [1.957]  \\ \hline
    $2{\rm D}\to2{\rm P}$      & 0.233 [0.240] & 0.874 [0.761] & 0.023 [0.025] & 0.440 [0.399] & 0.841 [0.597] & 1.104 [1.597] \\ \hline

    $3{\rm D}\to1{\rm P}$      & 0.247       & 0.792      & 0.027        & 0.431       & 0.645      &   1.039  \\ \hline
    $3{\rm D}\to2{\rm P}$      & 0.243       & 0.799      & 0.026        & 0.426       & 0.664      &   1.042  \\ \hline
    $3{\rm D}\to3{\rm P}$      & 0.230       & 0.880      & 0.022        & 0.434       & 0.877      &   1.085  \\ \hline

    $4{\rm D}\to1{\rm P}$      & 0.248       & 0.781      & 0.027        & 0.429       & 0.624      &   1.029  \\ \hline
    $4{\rm D}\to2{\rm P}$      & 0.246       & 0.781      & 0.027        & 0.425       & 0.627      &   1.027  \\ \hline
    $4{\rm D}\to3{\rm P}$      & 0.242       & 0.801      & 0.026        & 0.427       & 0.679      &   1.033   \\ \hline
    $4{\rm D}\to4{\rm P}$      & 0.226       & 0.953      & 0.022        & 0.471       & 0.999      &   1.185    \\ \hline

    $5{\rm D}\to1{\rm P}$      & 0.249       & 0.775      & 0.027        & 0.426       & 0.610      &   1.022   \\ \hline
    $5{\rm D}\to2{\rm P}$      & 0.247       & 0.773      & 0.027        & 0.423       & 0.608      &   1.019    \\ \hline
    $5{\rm D}\to3{\rm P}$      & 0.245       & 0.782      & 0.026        & 0.423       & 0.633      &   1.021    \\ \hline
    $5{\rm D}\to4{\rm P}$      & 0.240       & 0.825      & 0.025        & 0.437       & 0.707      &   1.069   \\ \hline
    $5{\rm D}\to5{\rm P}$      & 0.222       & 0.986      & 0.021        & 0.479       & 1.006      &   1.214   \\ \hline \hline
\end{tabular}
\end{center}
\caption{ The ratios among the radiative decay widths of the
$D$-wave   to the $P$-wave bottomonium states, where
$\widetilde{\Gamma}=\frac{\Gamma}{\Gamma(\Upsilon_3  \to\chi_{2}\gamma)}$, $\widetilde{\Gamma}(\Upsilon_3  \to
\chi_{2}\gamma)=\frac{\Gamma(\Upsilon_3  \to \chi_{2}\gamma)}{\Gamma(\Upsilon_3  \to \chi_{2}\gamma)}$=1, and
the values in the bracket come from the screened potential model \cite{LiChao0909} }
\end{table}

\section{Conclusion}
In this article, we extend our previous work on the radiative
transitions among the charmonium states to study the radiative
transitions among the bottomonium states in an systematic way  based
on the heavy quarkonium  effective theory, and make   predictions for
ratios among the radiative decay widths of a special multiplet to
another multiplet, where the unknown couple constants $\delta(m,n)$
are canceled out with each other. The predictions can be confronted
with the experimental data in the future at the Tevatron, KEK-B,
RHIC and LHCb.
\section*{Acknowledgment}
This  work is supported by National Natural Science Foundation of
China, Grant Number 11075053,  and the
Fundamental Research Funds for the Central Universities.

\end{document}